\documentclass[twocolumn,showpacs,preprintnumbers,amsmath,amssymb,prl]{revtex4}
\usepackage{hyperref}
\usepackage{aas_macros}
\usepackage{graphicx}
\usepackage{dcolumn}
\usepackage{bm}
\usepackage{array,booktabs}
\newcommand\VRule[1][\arrayrulewidth]{\vrule width #1}
\newcolumntype{C}[1]{>{\centering\let\newline\\\arraybackslash\hspace{0pt}}m{#1}}

\begin{document}

\preprint{APS/123-QED}

\title{Testing the Copernican principle with future radio-astronomy observations}

\author{H. L. Bester}
 \email{lbester@ska.ac.za}
\affiliation{SKA South Africa, 3rd Floor, The Park, Park Road, Pinelands, 7405 South Africa}
\affiliation{Department of Physics and Electronics, Rhodes University, PO Box 94, Grahamstown, 6140 South Africa}
\author{J. Larena}
 \email{j.larena@uct.ac.za}
 \affiliation{Department of Mathematics and Applied Mathematics, University of Cape Town, 7701 Rondebosch, South Africa}
 \author{N. T. Bishop}
 \email{n.bishop@ru.ac.za}
\affiliation{Department of Mathematics, Rhodes University, PO Box 94, Grahamstown, 6140 South Africa}

\date{\today}

\begin{abstract}
We use a direct observational approach to investigate the possibility of testing the Copernican principle with data from upcoming radio surveys. In particular we illustrate the importance of measuring derivatives transverse to the past light-cone when prior knowledge of the value of the cosmological constant is not available.  

\end{abstract}

\maketitle

\noindent \emph{Introduction} - The problem of reconstructing the large-scale cosmological metric directly from observations is a long-standing open problem which the observational cosmology programme \cite{Ellis1985} aims to solve. The direct observational approach invokes minimal assumptions about inaccessible parts of the Universe and constitutes a very robust test of the cosmological principle. The main difficulty with implementing this programme is obtaining data which do not presuppose any of the assumptions under scrutiny (i.e. model independent data). Indeed, since most astrophysical data gathering processes need a cosmological model at some stage or another, data which do not presuppose the cosmological principle at any stage whatsoever are surprisingly difficult to obtain. To simplify the problem, recent studies \cite{prd2010,prd2012,bester1,bester2} have focussed on the spherically symmetric case. Solutions to the general relativistic field equations for spherically symmetric dust universes are known as Lema\^itre-Tolman-Bondi (LTB) \cite{lemaitre33,tolman34,bondi47} or, if the cosmological constant is allowed to be non-zero, as $\Lambda$LTB models. These models necessarily violate the Copernican principle (CP) and, as the simplest generalisiation of $\Lambda$CDM models, are the simplest models which can be used to test for the presence of large scale (radial) inhomogeneities (see \cite{Bolejko:2016qku} and references therein for a recent review on inhomogeneous cosmology). In this letter we extend work initiated in \cite{prd2010,prd2012,bester1,bester2,Bester2016} by applying the \emph{Copernicus} algorithm \footnote{Code available at \url{https://github.com/landmanbester/Copernicus}.} to forecast data (\cite{Bester2016} should be consulted for a detailed review of our methodology).\\
The strongest evidence for the homogeneity and isotropy of the observed Universe come from observations of the cosmic microwave background (CMB) \cite{weiland2011,ade2013planck} and the growth of structure \cite{alam2015ApJS} particularly the constraints imposed by the presence of baryon acoustic oscillations (BAO) \cite{anderson2012}. The constraints from these observational features are usually derived by fitting parametrised models (i.e. solutions to the perturbed FLRW field equations) to the observed data. Although the success of these models lends strong credibility to concordance cosmology, there is an inherent circularity when using them to test the very assumptions on which they rely. \\
Previously, the kinematic Sunyaev-Zel’dovich \cite{kSZ1980} and integrated Sachs-Wolfe (ISW) \cite{ISW1967} effects have been used as probes for large scale radial inhomogeneity (e.g. see \cite{Zhang2011PhRvL,CliftonPhysRevD,Zibin:2014rfa,bolejo2014}) and were shown to significantly constrain such departures from FLRW. However, because of subtle but significant differences between the dynamics of multi-fluid cosmologies with different geometries \cite{Clarkson2011JCAP,Lim2013JCAP}, such studies need to be supplemented by independent methods such as the one employed in this letter. Note, in particular, that invoking a comoving description of non-interacting cold dark matter and radiation in a spherically symmetric space-time necessarily presupposes an FLRW metric (see \S A of \cite{Bester2016} for a proof). Because of its extremely low density at late times, radiation is unlikely to significantly alter the dynamics of dust after decoupling. However, the assumption that radiation is comoving with dust can only be justified along the central wordline of the observer (denoted $\mathcal{C}$) since this is the only maximally symmetric region of a spherically symmetric space-time. Therefore, strictly speaking, we would have to develop an analogous perturbative formalism around a spherically symmetric background in order to utilise the CMB to constrain the geometry of the Universe. Unfortunately, even for pure dust space-times, a perturbative analysis around spherical symmetry is far more complicated (see \cite{February2013qza}).\\ 
As a complementary analysis this work aims to investigate what constraints can be placed on the CP and the value of the cosmological constant while making as few as possible assumptions about the nature of the early Universe, thus focussing on late-time cosmology alone. The next section gives a very brief outline of our formalism, in particular how this goal can be achieved by measuring derivatives transverse to the current past light-cone (PLC0) of the observer. We then apply our algorithm to forecast data and discuss the possible model dependence of these data. Finally, we present a robust test for the CP which is able to quantify the degree to which observations allow departures from large scale radial inhomogeneities.\\
While our approach does not circumvent all of the above mentioned difficulties regarding the CMB, it depends on fundamentally different assumptions since we do not use the interaction of matter with the CMB directly. Instead we propose using data from upcoming intensity mapping surveys such as those planned for the Square Kilometre Array (SKA) \cite{SKA2016} and the Canadian Hydrogen Intensity Mapping Experiment (CHIME) \cite{CHIME2014} (a fairly complete overview of planned and upcoming intensity mapping surveys can be found in \cite{bull2014}). The formalism we present adds to the growing number of existing techniques (eg. \cite{Walters:2012ns,bolejo2014,Sapone:2014nna,2014MNRAS.438L...6V} and references therein) used to constrain the geometry of the Universe on large scales. \\

\begin{figure*}[t]
\includegraphics[width=0.8\textwidth]{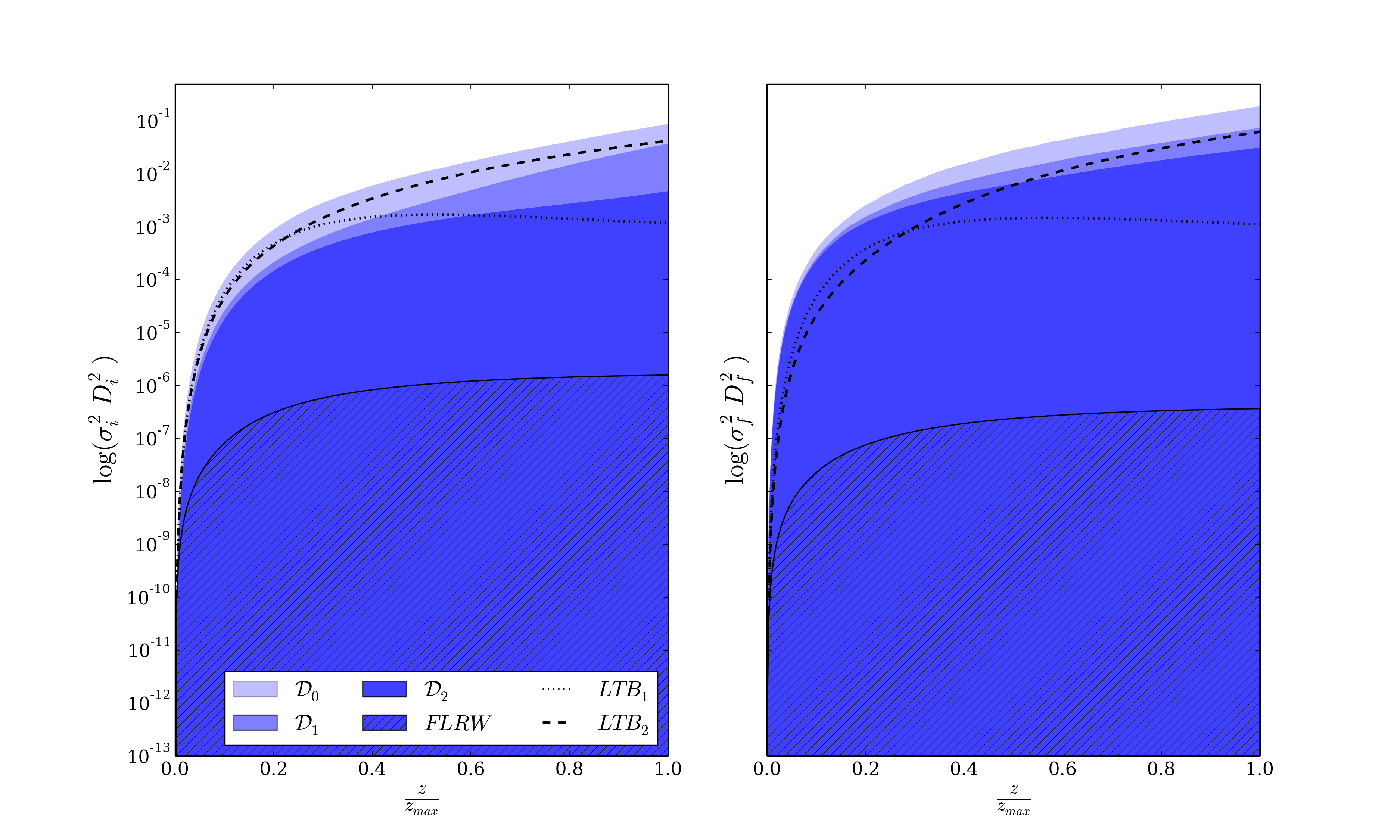}
\caption{The dimensionless quantity $\log(\sigma^2 D^2)$ on the current past light-cone (left) and on a past light-cone defined by $t \approx 10$Gyr (right) (note $\sigma^2$ is the fluid shear). The blue regions, from light to dark, correspond to the upper 2-$\sigma$ contours reconstructed from currently available data (i.e. simulation $\mathcal{D}_0$), forecast $D(z)$ and $\frac{dz}{dw}(z)$ data (i.e simulation $\mathcal{D}_1$) and finally all of the above including $H_\|(z)$ data from longitudinal BAO measurements (i.e. simulation $\mathcal{D}_2$, refer to Table \ref{tab:sims}). The hatched region is the same quantity corresponding to a perturbed FLRW model with a uv-cutoff of 100Mpc. For comparison we show two LTB models, one with a homogeneous bang time $t_B(r) \equiv 0$ (labelled $LTB_1$) and one without (labelled $LTB_2$).  \label{fig:sigmasq}}
\end{figure*}

\noindent \emph{Constraining $\Lambda$LTB with observations} - The spherically symmetric metric in observational coordinates $x^a = [w,v,\theta,\phi]$ can be written as (see \cite{Bester2016})
\begin{equation}
ds^2 = -A(w,v)dw^2 + 2dw dv + D(w,v)^2 d\Omega^2.
\label{ObsMet}
\end{equation} 
To solve the corresponding Einstein field equations we require the functional forms of $\rho(v)$ and $z(v)$ as well as the value of $\Lambda$. The inverse of the $z(v)$ relation required to interpret observables in terms of the null affine parameter $v$ follows from projecting the null geodesic equation along the direction of propagation of the ray and can be written as
\begin{equation}
v(z) = \int_0^{z_{\mbox{\tiny{max}}}} \frac{dz}{(1+z)^2 H_\|(z)},
\label{nuz}
\end{equation}
where $H_\|(z)$ is the observed expansion rate along the line of sight of the observer. Noting that this relation is one to one for positive $H_\|(z)$, we may also use it to find $z(v)$. Thus the combination of $\rho(z)$, $H_\|(z)$ and $\Lambda$ completely specifies a $\Lambda$LTB model. Moreover, any expanding $\Lambda$LTB model can be completely specified from these initial data. This is in stark contrast to the comoving approach to cosmology in which specific gauge choices and parametrisations limit model coverage from the outset. The fact that $\rho(z)$ and $H_\|(z)$ are in principle directly observable also makes it possible to inform the priors over the input functions directly by performing Gaussian process regression (GPR) \cite{rasmussen2006gp} on the data. This makes the observational approach preferable to the more conventional 1+3 approach to cosmology in which the input functions are specified on the current time slice of the observer. \\
An additional advantage of our approach is that we are able, using a numerical integration scheme (developed in \cite{bishop1997,prd1997} and adapted for cosmology in \cite{prd2010,prd2012}), to also reconstruct the geometry in the interior of the PLC0. This is important because homogeneity on the PLC0 does not necessarily translate into homogeneity in its interior. As we will see below, the data that we consider does not necessarily favour a more homogeneous universe in the past.\\  
Note that $\rho(z)$, $H_\|(z)$ and $\Lambda$ can be specified freely on any \emph{single} PLC. However, once fixed, these completely determine the evolution history of the Universe. Thus the value of the cosmological constant can, in principle, be inferred given data on any two distinct PLC's. Equivalently its value can be inferred by measuring derivatives transverse to the PLC0. For example, in observational coordinates, the redshift drift in $\Lambda$LTB models can be expressed as (see \cite{bester2,Bester2016})
\begin{equation}
\frac{dz}{dw} = \dot{z} + \frac{z'}{2}\left(A - \frac{1}{(1+z)^2}\right),
\label{DerTraPLC}
\end{equation}
where $w$ measures proper time along $\mathcal{C}$ and an overdot/prime refers to partial derivative with respect to $w$/$v$ respectively. Although similar expressions can be derived for arbitrary observables, the relative ease with which accurate redshift measurements can be made without presupposing a cosmological model suggests that redshift might be the most suitable observable with which to exploit this relation. \\
Below we use forecast $D(z)$ and $H_\|(z)$ data from \cite{bull2014} (for a Facility-type survey) and forecast $\frac{dz}{dw}(z)$ data from \cite{2014PhRvL.113d1303Y} to investigate the possibility of testing the CP with future radio observations. These data are simulated around a fiducial $\Lambda$CDM model defined by
\begin{equation}
\Omega_{m0} = 0.3, \quad \Omega_{\Lambda 0} = 0.7, \quad H_0 = 70 \frac{\mbox{km}}{\mbox{s Mpc}}.
\end{equation} 
We compare the results from forecast data to those derived from currently available distance modulus (equivalently angular diameter distance $D(z)$) data from the Union 2.1 compilation \cite{suzuki2012}, $H_\|(z)$ data from cosmic chronometers \cite{moresco2011,moresco2015} and a lower bound on the current age $t_0$ of the Universe \cite{sneden1996}. Note that without information on $t_0$ current data would be completely unable to constrain the value of $\Lambda$ \footnote{The structure of the field equations (see e.g. \cite{Bester2016}) establishes that, even if very precise $D(z)$, $H_\|(z)$ and $\rho(z)$ data were available, we still would not be able to constrain its value. The reason for this is that, on any single PLC, $H_\|(z)$ and $\rho(z)$ actually determine $D(z)$ independently of $\Lambda$. However, the value of $t_0$ depends on the history along the worldline of the observer $\mathcal{C}$ and therefore uses information from multiple PLC's. Since a $\Lambda$LTB model reduces to $\Lambda$CDM on $\mathcal{C}$ it does not matter that $t_0$ is determined assuming an FLRW geometry along $\mathcal{C}$.}. As it stands, since we only have a lower bound on $t_0$, current data only constrains $\Lambda$ from above. Also note that currently available $H_\|(z)$ data from cosmic chronometers have a weak model dependence imposed by the necessity to assume a homogeneous galaxy formation time \cite{deputter2012}. To investigate the importance of including $H_\|(z)$ data in the analysis we perform simulations with and without $H_\|(z)$ data. Table \ref{tab:sims} summarises the data used to evaluate the likelihood in each of the three simulations reported in Figure \ref{fig:sigmasq}.
\begin{table}[h]
\begin{tabular}{!{\VRule[1pt]}C{1.5cm}!{\VRule}C{1.5cm}!{\VRule}C{1.5cm}!{\VRule}C{1.5cm} !{\VRule}C{1.5cm}!{\VRule[1pt]}} 
\toprule[1pt]
\emph{Name} & $D(z)$ & $H_\|(z)$ & $t_0$ & $\frac{dz}{dw}(z)$ \\ \hline
$\mathcal{D}_0$ & \cite{suzuki2012} & \cite{moresco2011,moresco2015} & \cite{sneden1996} & - \\ \hline
$\mathcal{D}_1$ & \cite{suzuki2012,bull2014} & - & - & \cite{2014PhRvL.113d1303Y} \\ \hline
$\mathcal{D}_2$ & \cite{suzuki2012,bull2014} & \cite{moresco2011,moresco2015,bull2014} & \cite{sneden1996} & \cite{2014PhRvL.113d1303Y}  \\
\bottomrule[1pt]
\end{tabular}
\caption{Data sets used in the different simulations. The \emph{Name} column is an abbreviation for the simulation and corresponds to the caption entries in Figure \ref{fig:sigmasq}. The entries below each subsequent column indicates where the data were taken from. \label{tab:sims}}
\end{table}
The simulation $\mathcal{D}_0$ provides constraints from the currently available data that we consider. The simulation $\mathcal{D}_1$ uses only $D(z)$ and $\frac{dz}{dw}(z)$ to constrain the geometry of space-time, an idea that was investigated in \cite{Uzan2008PhRvL}. These two relations could be the most promising for the observational cosmology programme since they can both, in principle, be observed in a model independent way. Finally, the simulation $\mathcal{D}_2$ uses the full set of data considered in this work and should provide the tightest constraints. \\
For all simulations the prior on $H_\|(z)$ is set by performing GPR on cosmic chronometer data while the prior on $\rho(z)$ is set as explained in \cite{Bester2016}. To set the prior on $\Lambda$, we fix $H_\|(z)$ and $\rho(z)$ to their posterior GPR means and evaluate the likelihood on a $\Lambda$ grid consisting of 50 points between zero and an appropriate upper bound. We then fit a (possibly truncated) Gaussian distribution to the corresponding likelihood points and artificially inflate its standard deviation by 25\%. This makes it possible to use a preconditioned Crank-Nicolson sampler \cite{cotter2013} to perform inference on $\rho(z)$, $H_\|(z)$ and $\Lambda$. This sampler is designed to perform inference in function space and has an acceptance rate which is independent of the level discretisation used in the numerical integration scheme. We test for convergence of the chain by computing the potential scale reduction factor (PSRF) \cite{Gelman92} of each of the inputs \footnote{We compute an independent PSRF for each of the input functions at each value of $z$ in the domain.}. The sampler is terminated when the PSRF for each input is sufficiently close to one (typically $< 1.1$). The numerical error in the integration scheme is fixed to $10^{-5}$ throughout.\\
Note that all the data presented in Table \ref{tab:sims}, be it simulated or real, are, to some degree, biased towards homogeneous models. For real data this stems from the fact that the reported uncertainties in the cosmological observables (eg. angular diameter distance or expansion rate) are usually derived by fitting astrophysical models to the true observable quantities (eg. supernovae light-curves or the $4000\AA$ break). These astrophysical models contain unknown nuisance parameters which are usually inferred and marginalised over using a blind analysis \cite{conley2006,amanullah2010,suzuki2012} i.e. while simultaneously fitting a specific cosmology to the data. Clearly, if an FLRW cosmology is presupposed, the analysis will be biased towards homogeneous models. To obtain an unbiased estimate we would have to perform a blind analysis for all observables. Furthermore, the additional freedom present in $\Lambda$LTB models would require performing such a blind analysis simultaneously on multiple astrophysical observables. Considering how complicated it is to properly account for systematics in supernovae data alone (see \cite{Rubin2015ApJ} for example), such a simultaneous blind analysis would be highly non-trivial. The poor constraints (shown in Figure \ref{fig:sigmasq}) that can be derived from current data also suggest that such an analysis (which would likely only degrade the constraints) is premature, at least for the combination of data considered in this work. Thus, in all our simulations, we assume that the errors in the data are Gaussian and conditionally independent and use a simple $\chi^2$ distribution to evaluate the likelihood. A more realistic likelihood model will be investigated in future research. \\

\noindent \emph{Results and discussion - } Our results are summarised in Figure \ref{fig:sigmasq}. Since LTB void models are commonly used to test for the presence of large scale radial inhomogeneities, we also show the expected $\sigma^2 D^2$ from two parametrised LTB models. The first, labelled $LTB_1$, corresponds to an LTB model with a homogeneous bang time function (i.e. a simultaneous big bang $t_B(r) \equiv 0$) and a density profile parametrised as in \cite{gbh2008}. The second, labelled $LTB_2$, corresponds to an LTB model with both the bang time and the density profile parametrised as in \cite{gbh2008}. The specific parameters of the models shown in Figure \ref{fig:sigmasq} were determined by fitting them to the data used in simulation $\mathcal{D}_0$. Although the current data we consider are not able to rule out either of these models, we see that, when combined with future $D(z)$ and redshift drift data, we are able to rule out both with at least a 95\% confidence level on the PLC0. Note that, while the shear corresponding to $t_B(r) \equiv 0$ LTB models decreases in the past, this is not necessarily the case for arbitrary LTB models, and similarly for $\Lambda$LTB models. This is the most likely cause of the degradation in the constraints on the final PLC considered. \\
It is important to note that the aim of this work is not simply to rule out LTB void models as alternatives to dark energy but to test the CP in a more general setting. In order to say something meaningful about the validity of the CP we test whether the data are compatible with what is expected from perturbed FLRW cosmology. For this reason we also show the quantity $\langle \sigma^2 \rangle\bar{D}^2$ expected from a linearly perturbed FLRW model. Here $\bar{D}$ is the angular diameter distance in the background $\Lambda$CDM model and $\langle \sigma^2 \rangle$ is the expectation value of the scalar shear in a first order perturbed FLRW model, calculated in the longitudinal gauge. Since this quantity is divergent, and since we are only interested in the cosmological background where a smoothing scale is naturally expected, it is cut-off at scales smaller than $100$ Mpc for the purposes of illustration. Note that, to obtain quantities that make sense in both the FLRW and the $\Lambda$LTB cases, we compare the dimensionless quantity $\langle \sigma^2 \rangle\bar{D}^2$ in FLRW to $\sigma^2 D^2$ in the background $\Lambda$LTB models allowed by the data. If the constraints from the data can be shown to coincide with what is expected from perturbed FLRW, this would suggest that there is no reason to favour $\Lambda$LTB over $\Lambda$CDM as the model describing the background dynamics of the Universe. However, our results suggest that, even for our most optimistic simulation (viz. $\mathcal{D}_2$), it would be difficult to substantiate such a conclusion. What these simulations do tell us is that redshift drift would go a long way to providing accurate and independent information on the value of the cosmological constant. More specifically, with data of about the same quality as the forecasts of \cite{2014PhRvL.113d1303Y}, we would be able to determine the value of $\Omega_{\Lambda0}$ to within an uncertainty of about 0.1, independently of the underlying background. However, it should be kept in mind that these somewhat idealised simulations really provide maximal constraints. Indeed, we have neglected a number of factors that can significantly degrade the constraints, in particular accounting for uncertainties in the observed redshift. This is particularly relevant to redshift drift data since obtaining such data inevitably involves binning in redshift. Also, note that, the BAO analysis performed in \cite{bull2014} would have to be adapted for $\Lambda$LTB models; see \cite{February2013JCAP} for calculations of the BAO in LTB models. Such corrections would have to be taken into account for actual data. Alternatively, if we adopt the idea proposed in \cite{Uzan2008PhRvL} and use only $D(z)$ and $\frac{dz}{dw}(z)$ data, we could conceivably obtain $D(z)$ data of a similar quality from alternative sources such as the Dark Energy Survey \cite{DES2005}. \\
We have described a novel approach to testing the CP which uses very few assumptions about the nature of the early Universe. Our results suggest that data from upcoming HI surveys will significantly constrain the geometry of the Universe on large scales. Since continuum radio surveys such as the SKA are also expected to measure the radio dipole with high fidelity \cite{2015aska.confE..32S}, it will soon be possible to test the CP in a manner which is completely independent of the CMB. If it can be shown that such an analysis is consistent with the results of \cite{Zibin:2014rfa,Zhang2011PhRvL}, for example, there would really be very little doubt left as to the validity of the CP. However, as shown in Figure \ref{fig:sigmasq}, even very optimistic forecast $D(z)$, $H_\|(z)$ and $\frac{dz}{dw}(z)$ data will likely not be sufficient to constrain the geometry to within the necessary precision to validate the CP. This is largely because of the lack of constraints on $\rho(z)$. While galaxy surveys such as Euclid \cite{2011arXiv1110.3193L} are expected to yield accurate number densities of sources out to high redshift, the problem of converting number densities into the combined total energy density $\rho(z)$ of the Universe is very challenging, especially for inhomogeneous models \cite{iribarrem2013} (see \cite{iribarrem2014} however). This problem is closely related to the problem of using more realistic likelihood functions and will be investigated in future research. 

\noindent \emph{Acknowledgements - }
HLB is supported by the the South African Square Kilometre Array (SA SKA). JL and NTB are supported by the National Research Foundation (South Africa).

\bibliography{obsletter}

\end{document}